# Magnetic Control of Ferroelectric Polarization in a Self-formed Single Magnetoelectric Domain of Multiferroic $Ba_3NbFe_3Si_2O_{14}$

Nara Lee[1,2], Young Jai Choi[1,2*], and Sang-Wook Cheong[2]

[1]*Department of Physics and IPAP, Yonsei University, Seoul 120-749, Korea*

[2]*Rutgers Center for Emergent Materials and Department of Physics & Astronomy, Rutgers University, Piscataway, New Jersey, 08854, USA*

We have discovered strong magnetoelectric (ME) effects in the single chiral-helical magnetic state of single-crystalline langasite $Ba_3NbFe_3Si_2O_{14}$ that is crystallographically chiral. The ferroelectric polarization, predominantly aligned along the *a* axis below the Nèel temperature of ~27 K, changes in a highly non-linear fashion in the presence of in-plane magnetic fields (*H*) perpendicular to the *a* axis (*H*//*b**). This ME effect as well as smaller ME effects in other directions exhibit no poling dependence, suggesting the presence of a self-formed single ME domain. In addition, these ME effects accompany no-measurable hysteresis, which is crucial for many technological applications.

[*]phylove@yonsei.ac.kr

Multiferroics provide potentials for multifunctional device applications utilizing remarkable ME effects between electric and magnetic order parameters.[1,2] Especially, in magnetism-driven ferroelectrics, the ferroelectricity emerges when the magnetic order comes in with the broken inversion symmetry.[3-7] Since the ferroelectricity originates from the lattice relaxation via exchange strictions in the ordered magnetic state, significant cross-couplings are observed. In single phase multiferroics, the formation of multiple ME domains ordinarily prohibits achieving controllability of coupled parameters by applying electric and magnetic fields and, therefore, the complex poling process is required to generate a single ME domain. The multiferroic $Ba_3NbFe_3Si_2O_{14}$ (BNFSO) where a single magnetic domain is naturally formed has been suggested as a candidate for the demonstration of strong ME effects without the intricate poling process.[8-10] However, the significant ME effects in this compound have not yet been revealed.[11] The common magnetic hysteresis which occurs in the drastic change of dielectric properties through first order phase transitions can also be disadvantageous for many applications. Therefore, demonstrating the magnetic field induced change of dielectric properties without involving magnetic hysteresis can be of another use for practical potential of multiferroics.

BNFSO has drawn great attention because of its remarkable magnetic structure and multiferroicity.[8,12] It has a single crystallographically-chiral domain in which two chiral-helical magnetic states can be present, but only a single magnetic domain is experimentally observed.[8] As shown in Fig. 1(a), non-centrosymmetric BNFSO crystallizes in a rhombohedral structure with $Fe^{3+}$ ions (S=5/2) forming the isolated triangles in the *ab*-plane. It is characterized as the P321 crystallographic space group which is non-polar with a unit cell of *a*=*b*=0.8539 nm and *c*=0.52414 nm at room temperature. Among the five different exchange interactions ($J_1$ to $J_5$ in Fig. 1(a)), two inter-planar diagonal interactions, $J_3$ and $J_5$ are not equal due to the different bond lengths and angles. This leads to two possible magnetic domains regarding different spin helicity along the *c*-axis (shown on the bottom of Fig. 1(a)) and the triangular spin chirality in the plane (shown on the top of Fig. 1(a)). One sort of spin chirality which is selected by the sign of DM (Dzyaloshinsky-Moria) vector, $\vec{D}$, in the antisymmetric DM interaction $\vec{D} \cdot (\vec{S_i} \times \vec{S_j})$, fixes the spin helicity.[13] Thus, the magnetic ground state in zero magnetic field forms a single magnetic domain which was observed by the neutron diffraction[14,15] and electron spin resonance (ESR)[16] experiments.

The observation of a ferroelectric polarization in BNFSO has been reported and its magnitude reaches ~9 μC/m$^2$ at the lowest temperature.[9] However, the orientation of the polarization is theoretically arguable. A phenomenological model predicts that the helical spin and polarization axes are both parallel to the *c*-axis whereas Berry phase calculations give rise to a ferroelectric polarization pointing along the in-plane direction with an estimated polarization value of ~10 μC/m$^2$.[10,17] Moreover, the experimental control of ferroelectric properties by the external magnetic fields, which is the most fascinating signature of a magnetism-driven ferroelectric, has not yet been observed in this compound.

In our high-quality single crystal of BNFSO, we have confirmed that the ferroelectric polarization mostly points along the *a* axis and have observed a highly nonlinear ME effect. The result obtained without any measurable poling dependence suggests the presence of a self-formed single ME domain. The magnetic control of polarization involves no distinctive magnetic hysteresis.

Single crystals of BNFSO were grown by the floating zone method. DC magnetization, $M$, was obtained using a SQUID magnetometer (Quantum Design MPMS), specific heat was measured using the standard relaxation method with a Quantum Design PPMS, and dielectric constant, $\varepsilon'$, was measured in an LCR meter at $f$ =44 kHz. The temperature (magnetic field) dependence of electric polarization, $P$, was obtained by the integration of pyroelectric (magnetoelectric) current measured using an electrometer with the temperature (magnetic field) variation of 4 K/min (0.02 T/s).

Magnetic properties of our BNFSO crystals were measured along the three different orthogonal directions (*a*, *b**, and *c*). Fig. 1(b) shows magnetization curves in $H$=0.2 T. A clear anomaly occurs at $T_N$≈27 K with a substantial magnetic anisotropy. Below the $T_N$, the almost indistinguishable decreasing behavior of magnetizations along the *a* and *b** axes and the increase of magnetization along the *c* axis, indicate the easy-planar antiferromagnetic order of $Fe^{3+}$ spins. In Fig. 1(c), the isothermal magnetizations exhibit a small bending at low magnetic fields and then increases linearly up to 7 T. They also reveal the slight magnetic anisotropy, consistent with the previous reports.[18,19]

In our thorough investigation of ferroelectric properties along the three orthogonal axes, the clear evidence of the multiferroicity of BNFSO is obtained from the onset of ferroelectric polarizations at $T_N$=27 K under zero magnetic field. As shown in Fig. 2(a), the ferroelectric polarizations emerge for all three different orientations below $T_N$. The polarization along the $a$-axis ($P_a$) appears with the magnitude of $P_a \approx 7$ μC/m$^2$ and shows almost linear increase as the temperature decreases. $P_c$ and $P_{b*}$ are only about 1 μC/m$^2$ at the lowest temperature. In Fig. 2(b), the dielectric constant along the $a$ axis ($\varepsilon_a$) exhibits the weak anomalous feature at $T_N$ consistent with the small magnitude of electric polarization, while the specific heat reveals the sharp feature at $T_N$. Our observation of $P_a$ is consistent with the THz-spectrum experimental result and Ginzburg-Landau free energy analysis. It is also compatible with the recent structural study which reveals that the non-polar P321 symmetry is lowered to polar C2 symmetry at low temperature, that is, $Fe^{3+}$ triangles are no longer equilateral and the three fold symmetry is broken.[20]

We emphasize that the observation of a ferroelectric polarization without any poling dependence is unique. In magnetism-driven ferroelectrics, the magnitude of a polarization is quite small compared to that in the conventional ferroelectrics. Therefore, temperature and magnetic field dependences of polarizations are attained using the pyroelectric and ME current measurements, respectively. Cooling from a temperature above a multiferroic transition under the external electric and/or magnetic fields creates a single ME domain which gives rise to the fully saturated magnitude of ferroelectric polarization. In the case of BNFSO, no poling dependence is detected in the magnitude of polarization. This suggests that the single ME domain is naturally formed at the multiferroic transition temperature.

Figure 3 shows the isothermal magnitude change of ferroelectric polarizations with sweeping the applied magnetic fields from +9 T to −9 T, and from −9 T to +9 T at 2 K. The major changes were observed in $H_{b*}$ dependence of $\Delta P_a$ (Fig. 3(a)) and $H_a$ dependence of $\Delta P_{b*}$ (Fig. 3(b)). With increasing the magnetic fields, the magnitude change of polarizations persists up to the maximum measuring field of 9 T. The increasing behavior follows the parabolic relation up to 4 T and becomes linear above 4 T. However, this nonlinear to linear behavior is not evidenced in the $H$ dependence of magnetization. Upon decreasing the magnetic fields, no distinct magnetic hysteresis was observed in our isothermal magnetic field dependence of ferroelectric polarizations. Most examples of tunable polarizations and dielectric constants with external magnetic fields in multiferroics involve a large field hysteresis resulting from first-order magnetic transitions,[21,22] which can be disadvantageous for many technological device applications. However, the magnetic hysteresis in the magnetic field driven variation of polarizations in BNFSO turns out to be negligible. The hysteretic difference is quantified as the polarization change between the ramping-up and down magnetic fields normalized by the polarization difference in $H$=0 and 9 T. The estimated differences are less than 1.26 % for $\Delta P_a(H_{b*})$ and 1.84 % for $\Delta P_{b*}(H_a)$.

The temperature dependences of $P_a$ and $P_{b*}$ in various magnitudes of $H_{b*}$ and $H_a$, respectively, were also measured. In the figures shown ahead, the absolute magnitudes of polarizations (Fig. 2(a)) and their variations in magnetic fields (Fig. 3) are only presented. But the directions of $P_a$ and $P_{b*}$ are indeed reversed in $H_{b*}$ and $H_a$, respectively, as shown in Fig. 4. In Fig. 4(a), $P_a$ below $H_{b*}$=6 T shows the negative values in all temperature regime below $T_N$. In $H_{b*}$=7 T, the negative $P_a$ changes to a small positive value below 10 K, and above $H_{b*}$=7 T, $P_a$ appears to be fully reversed. In Fig. 4(b), small negative value of $P_{b*}$ in $H_a$ =0 changes little up to 3 T, but becomes positive and increases significantly with increasing $H_a$ above 3 T. Since the exact magnetic structure in magnetic fields has not yet been constructed, it is unclear why the electric polarizations are opposed by external magnetic fields. Therefore further investigations to verify the magnetic structure and possible mechanism for the polarization reversals are certainly desirable.

In summary, our results disclose the intricate magnetoelectric behavior in $Ba_3NbFe_3Si_2O_{14}$. The ferroelectric polarization predominantly directing along the *a* axis accompanies the natural development of the single magnetoelectric domain in this multiferroic. $P_a$ is opposed by applying $H_{b*}$ and the tiny $P_{b*}$ is reversed significantly under $H_a$. The magnetic field driven change of polarizations involving no poling dependence and little magnetic hysteresis provides potentially applicable functionalities in multiferroics.

**Figure Captions**

**FIG. 1** (Color online). (a) (Top) Planar crystallographic structure of $Ba_3NbFe_3Si_2O_{14}$. Three orthogonal crystallographic orientations are denoted as $a$, $b^*$, and $c$. Blue arrows indicate $Fe^{3+}$ spins with one fixed triangular chirality. $J_1$ and $J_2$ are the nearest intra- and inter-triangular couplings, respectively, in the plane. (Bottom) Two adjacent triangles of $Fe^{3+}$ ions along the $c$-axis. Spin arrangement in each triangle is displayed separately and shows spin helicity along the $c$-axis. $J_3$ to $J_5$ are the inter-planar interactions between adjacent triangles along the $c$-axis. Two inter-diagonal interactions, $J_3$ and $J_5$ are asymmetric. (b) Temperature dependent magnetic susceptibility measured in $H$=0.2 T along the $a$, $b^*$, and $c$ axes. Dotted line indicates the Nèel temperature of ~27 K. (c) Isothermal magnetization with both ramping up and down measurements along the $a$, $b^*$, and $c$ axes at 2 K.

**FIG. 2** (Color online). (b) Temperature dependence of ferroelectric polarizations in absolute magnitudes along the $a$, $b^*$, and $c$ axes in zero magnetic field. Polarization emerges mainly along the $a$ axis. (a) Temperature dependence of dielectric constant along the $a$-axis and specific heat divided by temperature in $H$ =0 T.

**FIG. 3** (Color online). Variation of ferroelectric polarizations in magnetic fields at 2 K. No distinct magnetic hysteresis was observed between ramping up and down curves. (a) $\Delta P_a (H_{b^*})$ (blue) and $\Delta P_a (H_c)$ (green). (b) $\Delta P_{b^*} (H_a)$ (red) and $\Delta P_{b^*} (H_c)$ (green). (c) $\Delta P_c (H_{b^*})$ (blue). The most pronounced changes of polarizations occur in $P_a (H_{b^*})$ and $P_{b^*} (H_a)$.

**FIG. 4** (Color online). Temperature dependence of ferroelectric polarizations at various magnetic fields. (a) $P_a$ in $H_{b^*}$ = 0, 4, 6, 7, 8, and 9 T. $P_a$ is reversed in partial temperature regime at 7 T and is totally reversed above 8 T. (b) $P_{b^*}$ in $H_a$ = 0, 3, 4, 6, 8, and 9 T. Small negative $P_{b^*}$ is entirely opposed above $H_a$ = 4 T.

**Acknowledgements**
We thank M. Mostovoy and J. H. Han for insightful discussions. The work at Yonsei University was supported by the NRF Grant (NRF-2012M2B2A4029730 and NRF-2013R1A1A2058155) and BK21 Plus project. The work at Rutgers University was supported by the DOE under Grant No. DE-FG02-07ER46382.

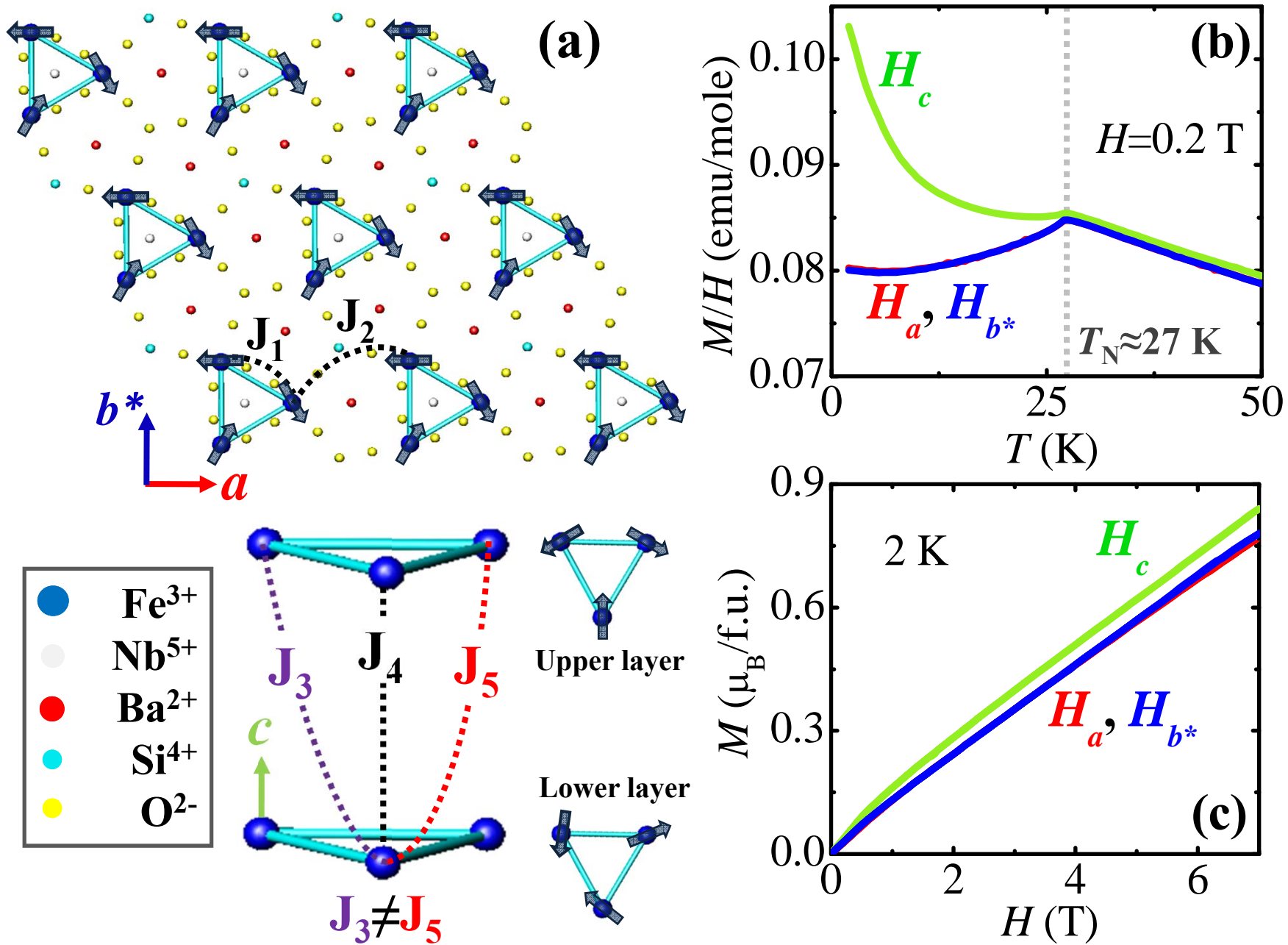

Figure 1 N. Lee *et al.*

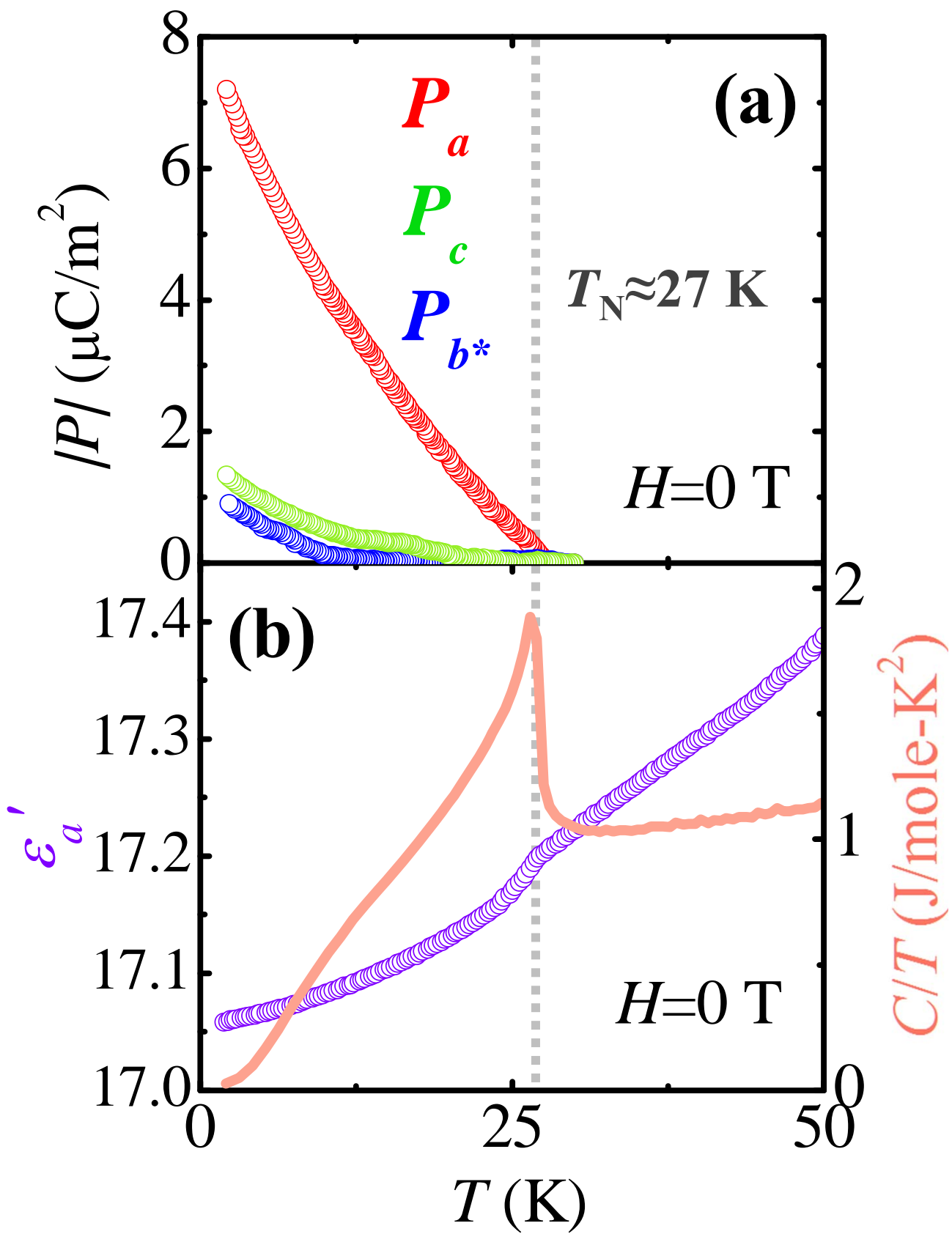

Fig. 2 N. Lee *et al.*

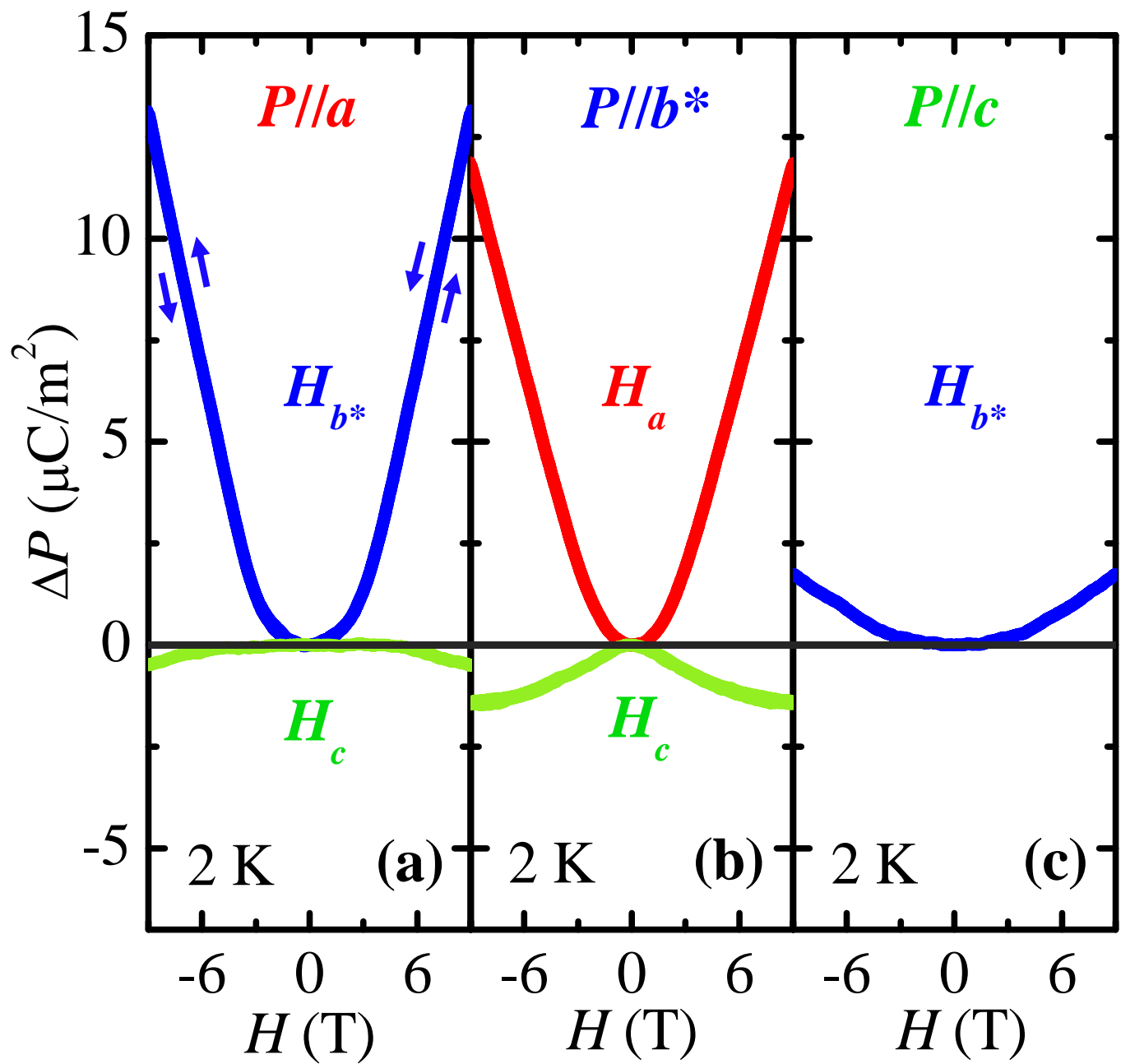

Fig. 3 N. Lee *et al.*

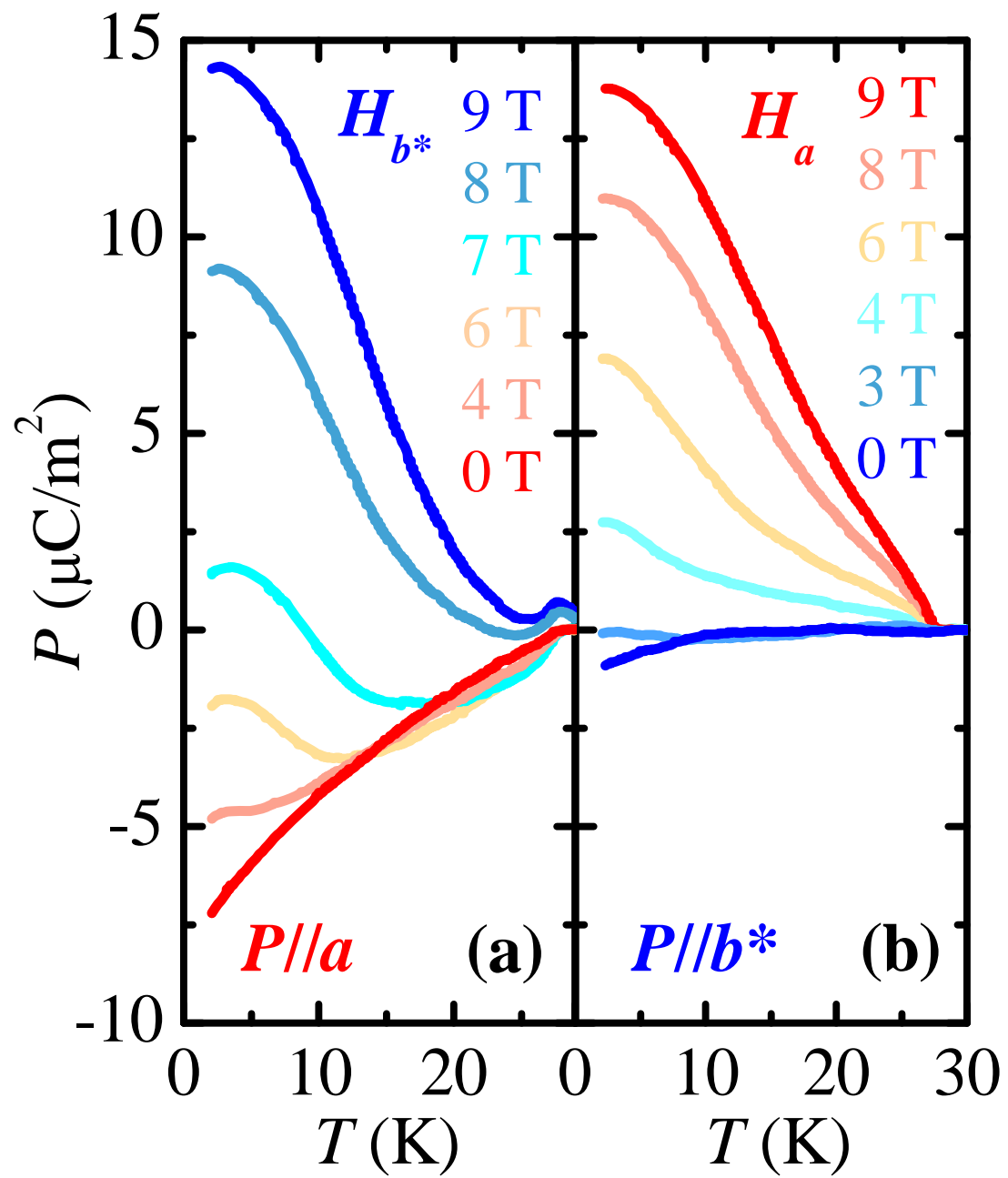

Fig. 4 N. Lee *et al.*